\documentclass[twocolumn,aps,pra,groupedaddress]{revtex4}

\usepackage{graphicx}
\begin{document}

\title
{
Distinguishing genuine entangled two--photon--polarization states\\
from independently generated pairs of entangled photons
}

\author{Kenji Tsujino}
 \altaffiliation[Electronic address: ]{tsujino@es.hokudai.ac.jp}
\author{Holger F. Hofmann}
 \altaffiliation[Electronic address: ]{h.hofmann@osa.org}
 \altaffiliation[Also at ]{PRESTO project, Japan Science and Technology 
 Agency.}
\author{Shigeki Takeuchi}
 \altaffiliation[Electronic address: ]{takeuchi@es.hokudai.ac.jp}
 \altaffiliation[Also at ]{CREST project, Japan Science and Technology 
 Agency.}
\author{Keiji Sasaki}
 \altaffiliation[Electronic address: ]{sasaki@es.hokudai.ac.jp}
\affiliation{%
	Research Institute for Electronic Science, Hokkaido University, 
	Sapporo 060--0812, Japan
}%

\date{\today}
%

\begin{abstract}
A scheme to distinguish {\it entangled two--photon--polarization 
states} ($ETP$) from two independent {\it entangled one--photon--polarization 
states} ($EOP$) is proposed. Using this scheme, the experimental generation of 
$ETP$ by parametric down--conversion is confirmed 
through the anti--correlations between three orthogonal 
two--photon--polarization states. The estimated fraction of $ETP$ among 
the correlated 
photon pairs is 37$\%$ in the present experimental setup. 
\end{abstract}

\maketitle
%
Entanglement is one of the key features of quantum theory, and the 
generation of {\it entangled one--photon--polarization states} ($EOP$)
by parametric down--conversion 
has been the focus of much experimental research \cite{Mandel88, Kwiat95, 
Kwiat99}. Such entangled photon pairs have also 
been employed in several experiments on quantum information processing 
and quantum communication \cite{ZeilingerBook}. At present, 
there appear to be two ways to expand the concept of entangled photons: 
increase the number of local systems, or increase the number of photons 
in the local system. In the former case of multi--party entanglement, 
single photons are distributed to multiple parties. This case has 
been realized 
through the generation of three--party entangled states such as GHZ states 
\cite{GHZ_1, GHZ_2, GHZ_3}, 
and various four--party entangled states \cite{FourPhotonEntangle}. 
In the latter case, $n$ photons conformed to the same spatiotemporal 
mode are distributed to each party. The first step in this direction is 
the generation of {\it entangled two--photon--polarization states} ($ETP$).

The two--photon--polarization states can be expressed using a basis 
of three orthogonal states \cite{ThreeStates}, exploiting the 
indistinguishability or bosonic nature of the two 
photons. Thus, two--photon--polarization states may be used as physical 
representations of three--level 
systems. Howell {\it et al}. \cite{Howell02} recently reported the 
violation of a Bell's inequality for 
spin--1 systems (three--level systems) proposed by Gisin and 
Peres \cite{GisinPeres92} using $ETP$. 
The quantum mechanical prediction of the maximum value of the conclusion 
for the $ETP$ is 2.55, whereas the classically predicted maximum is 2. 
Howell {\it et al.} experimentally confirmed the violation of this inequality, 
obtaining an experimental value of 2.27 $\pm$ 0.02, attributing the result 
to $ETP$.
However, the present authors have found that 
two independent EOPs, which may have been generated in 
the same experimental setup, can also violate
the Bell's inequality used by Howell {\it et al.},
with a value of up to 2.41 for the correlations. 
This is possible because the correlations in Bell's 
inequalities are not a measure of indistinguishability, 
but rather of non--locality.
Thus, the violation of a Bell's inequality reported
by Howell {\it et al}. in \cite{Howell02}
does not require the generation of $ETP$ and 
an alternative method for obtaining 
direct evidence of the generation of $ETP$ is desirable.

In this letter, we therefore propose a novel method to distinguish 
$ETP$ from two independent $EOP$s. In the proposed method, 
the orthogonality of three two--photon--polarization basis 
states is 
checked via the correlations between the two local systems. The experimental 
generation of $ETP$ using parametric down--conversion can then be evaluated 
using the proposed method. The results clealy reveal an anti--correlation 
between the corresponding basis states, indicating the successful 
generation of $ETP$.

$EOP$ can be generated by 
pulsed type--II parametric down--conversion \cite{Kwiat95}. When one pair 
of photons is emitted into 
two modes A and B, the quantum state of the pair 
is ideally given by
%
\begin{equation}
|EOP\rangle = (|H\rangle_A |V\rangle_B - |V\rangle_A |H\rangle_B)/\sqrt{2},
\end{equation}
%
where $H$ and $V$ represent horizontal and vertical polarization. This 
state has been used in many experiments on quantum information 
using photons \cite{ZeilingerBook}. 
However, in general, multi--pair states are also generated in 
higher--order processes \cite{LamasLinares01}. When two pairs are 
emitted simultaneously, the quantum state of the pairs 
is ideally given by
%
\begin{eqnarray}
|ETP\rangle 
 & = & \frac{1}{\sqrt{3}}(|HH\rangle_A |VV\rangle_B \nonumber \\
 & - & |HV\rangle_A |HV\rangle_B + |VV\rangle_A |HH\rangle_B),
\end{eqnarray}
%
where $|HH\rangle$, $|HV\rangle$ and $|VV\rangle$ are the three 
orthogonal basis states of two--photon--polarization states, corresponding 
to two $H$--polarized photons, one photon $H$--polarized and 
$V$--polarized each, and two $V$--polarized photons, respectively. Note 
that these states differ from the simple direct products of two 
photons, $|H\rangle_1 \otimes |V\rangle_2$, where 1 and 2 denote 
spatially or temporally independent modes.

The proposed method for distinguishing $ETP$ from 
two independent $EOP$s is as follows. 
The basis states of Eq.~(2) can be transformed into three 
orthogonal unpolarized photon states, i.e.,
%
\begin{eqnarray}
|ETP\rangle 
 & = & \frac{-1}{\sqrt{3}}(|HV\rangle_A |HV\rangle_B \nonumber \\
 & + & |PM\rangle_A |PM\rangle_B + |RL\rangle_A |RL\rangle_B),
\end{eqnarray}
%
where
%
\begin{equation}
|PM\rangle = \frac{1}{\sqrt{2}}(|HH\rangle - |VV\rangle),
\end{equation}
\begin{equation}
|RL\rangle = \frac{i}{\sqrt{2}}(|HH\rangle + |VV\rangle).
\end{equation}
%
Here, $P$, $M$, $R$ and $L$ represent plus--diagonal ($|P\rangle=
[|H\rangle+|V\rangle]/\sqrt{2}$), minus--diagonal ($|M\rangle=
[|H\rangle-|V\rangle]/\sqrt{2}$), right--circular ($|R\rangle=
[|H\rangle+i|V\rangle]/\sqrt{2}$) and left--circular ($|L\rangle=
[|H\rangle-i|V\rangle]/\sqrt{2}$) polarizations. Thus, $|PM\rangle$ 
represents a $P$--polarized photon and an $M$--polarized photon 
generated in the same mode, and $|RL\rangle$ denotes an $R$--polarized 
photon and an $L$--polarized photon. Note that these three states 
also form a set of orthogonal basis states \cite{ThreeStates}. 

Consider the case that one of the two photons of Eq.~(3), say 
that in path A, is detected in the $H$--polarized state, and the 
other photon also in path A is detected in the $V$--polarized state. 
In this situation, the two--photon--polarization states in mode A 
would be projected into 
the $|HV\rangle$ state, and the two--photon--polarization state 
in mode B would also be 
projected automatically into the $|HV\rangle$ state due to the 
entanglement between paths A and B. 
Consequently, the probability of detecting two photons in mode B 
in $|RL\rangle$ 
or $|PM\rangle$ states will be zero. Therefore, in actual experiments, 
a four--fold coincidence event in which the two photons in either mode 
are detected with different polarization ($H$ and $V$ in mode A, $P$ ($R$) 
and $M$ ($L$) in mode B) will never occur for pure state $ETP$. 
In the following, the number of coincidence events given the same 
measurement bases is 
defined as $C_{\parallel}$, and that for different measurement 
bases is defined as $C_{\perp}$. In the case of ideal pure state $ETP$, 
the ratio $r = C_{\perp}/C_{\parallel}$ is obviously 0. 

Next, consider the case in which two independent $EOP$s described 
by Eq.~(1) are emitted into spatially or temporally separable modes. 
The state of these two pairs is given by
%
\begin{eqnarray}
\lefteqn{|EOP \otimes EOP \rangle}\nonumber \\ 
& = & \frac{1}{\sqrt{2}}
(|H\rangle_{A1} |V\rangle_{B1} - |V\rangle_{A1} |H\rangle_{B1}) \nonumber\\
& \otimes & \frac{1}{\sqrt{2}}
(|H\rangle_{A2} |V\rangle_{B2} - |V\rangle_{A2} |H\rangle_{B2}),
\end{eqnarray}
%
where modes A1 and A2 (B1 and B2) are in the same optical path A (B) 
but are spatially or temporally distinguishable. For example, when two 
photons of Eq.~(6) in path A are measured in the $H/V$ basis, $|V\rangle_{A1}
\otimes|H\rangle_{A2}$, the state in path B will be $|H\rangle_{B1}
\otimes|V\rangle_{B2}$ due to the polarization entanglement. Suppose the 
two photons in modes B1 and B2 are detected in the $R/L$ basis, i.e. 
$|R\rangle_{B1}\otimes|L\rangle_{B2}$ or 
$|L\rangle_{B1}\otimes|R\rangle_{B2}$. 
Since the two photons are independent, the conditional probability 
for detecting two photons with polarization $|R\rangle$ and $|L\rangle$ when 
the other two photons in path A are detected with polarization $|H\rangle$ 
and $|V\rangle$ is $|{}_{B1}\langle R|H \rangle_{B1} 
{}_{B2}\langle L|V \rangle_{B2}|^2 + |{}_{B1}\langle L|H \rangle_{B1}
{}_{B2}\langle R|V \rangle_{B2}|^2 = 1/2$. Therefore, the ratio $r$ 
for two independent $EOP$s should be 1/2 even in the ideal case of 
pure state emission.

Since the correlations between the polarizations in A and B
should be maximal for the ideal pure state, it is reasonable 
to assume that r will be greater or equal to 1/2 in the more
general case of mixed state $EOP$s. Specifically, depolarization
effects due to experimental imperfections in the alignment of
the optical setup will typically increase the value of $r$.
For realistic mixed state $EOP$s, we can therefore assume that 
$r\geq 1/2$ \cite{footnote}.

These results suggest that the $ETP$ can be distinguished 
from two independent $EOP$s by measuring 
the correlation of the three orthogonal polarization basis states. 
According to the results given above, the following condition indicates
the presence of $ETP$ among emitted photons,
%
\begin{equation}
r<\frac{1}{2}.
\end{equation}
%
\begin{figure}[htbp]
	\includegraphics{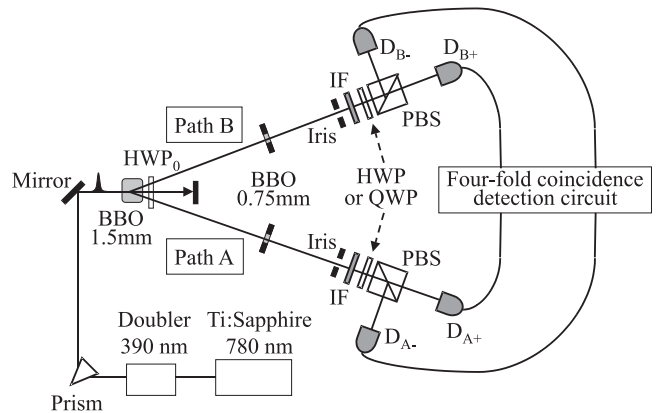}
	\caption{
	%
	Experimental setup for generation of entangled photon pairs 
	and measurement of polarization correlation between paths A 
	and B. Entangled pairs are generated in a beta--barium borate 
	(BBO) crystal by pulsed parametric down--conversion. $|HV\rangle$, 
	$|RL\rangle$ and $|PM\rangle$ states are measured using 
	combinations of a polarizing beam splitter (PBS), a quarter--wave 
	plate (QWP) at 45${}^{\circ}$ and a half--wave plate (HWP) at 
	22.5${}^{\circ}$ in each path. The events are counted when 
	the photon detectors $D_{A+}$, $D_{A-}$, $D_{B+}$ and $D_{B-}$ 
	are fired simultaneously.
	}
	\label{}
\end{figure}

The proposed method was used to evaulate the generation of $ETP$ 
using the experimental setup shown in Fig.~1. The pump laser was 
a 100 fs--pulsed, frequency--doubled Ti:sapphire laser (82 MHz repetition 
rate, $\lambda$ = 390 nm). The pump laser was focused onto a beta--barium 
borate (BBO, 1.5 mm) crystal using a convex lens ($f$ = 45 cm) in order 
to collect emitted photons efficiently \cite{Kurtsiefer01}. The pump 
laser pulses were incident on the BBO crystal, which was cut 
appropriately for type--II phase matching. The Kwiat '95 source condition 
\cite{Kwiat95} was adopted for the generation of $EOP$.
A half--wave plate (HWP${}_0$) and 0.75 mm BBO crystal were 
inserted in each path for compensation of spatiotemporal walk--off. 
In each of the paths, the generated photons pass through an 
interference filter IF (bandwidth 3.6 nm, centered at 780 nm) and a 
5 mm iris set 1 m from the crystal. 

The three basis states of two--photon--polarization were measured in 
each path using a half--wave plate (HWP), a quarter--wave plate (QWP), 
a polarizing beam splitter (PBS) and a pair of photon detectors 
(SPCM AQ/AQR series, PerkinElmer). For example, in path A, 
the PBS was inserted, and the HWP and QWP removed in order to measure the 
coincidence events associated with the basis state $|HV\rangle$ in path A. 
For measuring $|RL\rangle$, a PBS and QWP rotated by 45${}^\circ$ 
from the vertical were inserted. Similarly, 
a combination of a HWP rotated by 22.5${}^\circ$ and a PBS was used 
to detect the $|PM\rangle$ states. Continuous transfer of the 
measurement basis from one to the next among the three orthogonal 
basis states of Eq.~(3) was achieved by rotating the wave plate 
continuously. In order to measure correlation between path A and B 
in terms of polarization, four--fold coincidence events were counted 
using a electronic circuit and photon counter 
(SR400, Stanford Research Systems). To evaluate the experimental 
setup itself, two--fold coincidence events between all pairs of 
the four photon--detectors setup on paths A and B were measured. 
Coincidence fringes were observed according to the state $|EOP\rangle$ 
given by Eq.~(1), with fringe visibility of more than 90$\%$ for all 
combinations of detector pairs. 

\begin{figure}[htbp]
	\includegraphics{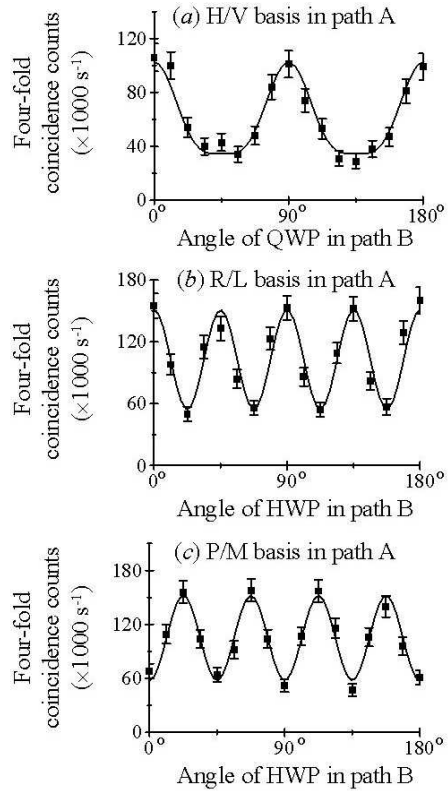}
	\caption{
	%
	Four--fold coincidence counts for ($a$) fixed $H/V$ polarization 
	measurement basis in path A while rotating QWP in path B, 
	($b$) fixed $R/L$ basis in path A while rotating HWP in path B, 
	and ($c$) fixed $P/M$ basis in path A while rotating HWP in path B. 
	Note that in Fig.~2($b$), a QWP at 45${}^{\circ}$ was inserted in path B. 
	Maximum and minimum counts were obtained with similar and 
	different polarization settings in each path, respectively. 
	The error bars represent the statistical errors estimated for a 
	Poisson distribution of the measurement results. 
	Pumping power was ($a$) 380 mW, ($b$) 410 mW 
	and ($c$) 420 mW. Solid lines represent the fitted theoretical prediction.
	}
	\label{}
\end{figure}
%
Fig.~2 shows an example of the experimental data. The error bars 
represent the statistical errors estimated for a Poisson 
distribution of the measurement results. Fig.~2($a$) shows the 
results for four--fold coincidence events while rotating the QWP in 
path B, with the polarization in path A fixed at the $H/V$ basis. The 
four--fold coincidence counts were maximal at QWP angles of 0${}^\circ$, 
90${}^\circ$ and 180${}^\circ$, corresponding to the $H/V$ basis 
measurement in path B, whereas minimal counts were recorded at 45${}^\circ$
 and 135${}^\circ$ ($R/L$ basis). Taking the averages of these values for 
$C_{\parallel}$ and $C_{\perp}$ over five experiments, the ratio $r$ was 
obtained as 0.36 $\pm$ 0.06. As this result satisfies the condition given in 
Eq.~(7), we find that this correlation indicates the successful 
generation of $ETP$.

Similarly, Fig.~2($b$) (Fig.~2($c$)) shows the results of 
four--fold coincidence 
counts with the HWP rotated in path B and $R/L$ ($P/M$) basis 
measurement setup in path A. In this case, the polarization basis was 
transformed from the $R/L$ ($H/V$) basis to the $P/M$ ($P/M$) basis, allowing 
different and similar polarization settings between paths A and B 
to be measured. Note that in order to transform 
from the $R/L$ basis to the $P/M$ basis, we set 
a QWP rotated by 45${}^\circ$ before the HWP in path B.
The obtained values of the ratio $r$ are 0.38 $\pm$ 0.05 
for $R/L$, and 0.36 $\pm$ 0.02 for $P/M$, which also satisfy 
condition (7) and are consistent with the result in Fig.~2($a$). 
Figs.~2($a$)--($c$) show a correlation 
between the same measurement outcomes 
($|HV\rangle$, $|RL\rangle$ and $|PM\rangle$), 
and anti--correlation between different measurement outcomes 
($|HV\rangle$--$|RL\rangle$, $|RL\rangle$--$|PM\rangle$ and 
$|PM\rangle$--$|HV\rangle$). Based on these results, 
specifically the correlation/anti--correlation ratios $r$ satisfying Eq.~(7), 
it appears that the present system successfully generated $ETP$. 

Perfect anti--correlation, which should have been obtained for 
ideal pure state $ETP$, was not achieved in these experiments. One of 
the reasons for this may have been the presence of $EOP$ due to 
a lack of coherence between the two emitted pairs. 
In the present experimental setup, temporal and spatial coherence 
was implemented using irises and interference filters. This may 
not have been sufficient to obtain single mode.

The ratio of $ETP$ among the generated states can be 
estimated by calculating the four--fold coincidence rates for 
a mixture of pure state $ETP$ (Eq.~(3)) and two independent pure state $EOP$s 
(Eq.~(6)). If we define $\alpha$ as the 
probability of the generation of pure state $ETP$ given by Eq.~(3) 
and $\beta$ as the probability of generating two independent 
pure state $EOP$s given by Eq.~(6), the four--fold coincidence rates are given by 
\begin{equation}
C_0 \left[ 
\frac{\alpha}{3}\cos^4(2\theta_{\lambda/4})  
+ \frac{\beta}{2} \left( \frac{\cos^4(2\theta_{\lambda/4})+1}{2} \right)
\right],
\end{equation}
\begin{equation}
C_0 \left[ 
\frac{\alpha}{3}\cos^2(4\theta_{\lambda/2})  
+ \frac{\beta}{2} \left( \frac{\cos^2(4\theta_{\lambda/2})+1}{2} \right)
\right],
\end{equation}
and
\begin{equation}
C_0 \left[ 
\frac{\alpha}{3}\sin^2(4\theta_{\lambda/2})  
+ \frac{\beta}{2} \left( \frac{\sin^2(4\theta_{\lambda/2})+1}{2} \right)
\right],
\end{equation}
%
where, Eqs.~(8)--(10) correspond to Fig.~2($a$), 2($b$) and 2($c$), 
respectively. $C_0$ is the total rate of four--fold coincidence counts and 
$\theta_{\lambda/4}$ and 
$\theta_{\lambda/2}$ are the angle of the QWP and HWP in path B, 
respectively (solid lines in Figs.~2($a$)--($c$)). Defining $r= 
C_{\perp}/C_{\parallel}$ as the ratio of the minima and maxima 
in Eqs.~(8)--(10), we obtain the relation between 
$\alpha$ and $r$ as follows. 
%
\begin{equation}
\alpha = \frac{1-2r}{1-2r/3}.
\end{equation}
%
Therefore, for the ratio of $r$ = 0.36 obtained in our experiments, 
we obtain an $ETP$ fraction of $\alpha$ = 0.37. 

The analysis given above is based on the assumption that 
the polarization states are the maximally coherent pure 
states given by Eq.~(3) and (6) and 
that the only source of coincidences between $H/V$ and $R/L$ is 
the multi-mode component given by the $EOP$ state in Eq.~(6). 
However, decoherence resulting
in mixed state outputs may also contribute to such coincidences.

In the present setup, the visibility of the coincidence counts observed 
for individual $EOP$ emission was about 0.9 at a count 
rate of 3000 s${}^{-1}$. 
In principle, complete quantum tomography would 
be needed to identify the effects of this decoherence on the 
two--photon--polarization states. In order to obtain a rough 
estimate of the effects of decoherence, it may be useful to consider the 
coincidence counts caused by white noise in the mixed
state $EOP$ emission. In this case, 
the rate of coincidence counts is equal to $1/4$ of the total rate of 
two pair coincidences, regardless of the polarizations measured. 
A white noise fraction of $\gamma$ added to the total density matrix 
therefore adds a constant background coincidence rate of 
$C_{0}\gamma/4$ 
to the polarization dependence given by Eqs.~(8)--(10).

Since the visibility of 0.9 observed in our experiment suggests a noise 
background of about 0.1 for one pair, we estimate that the two 
pair noise may be about $\gamma=0.2$. With this assumption, 
the single mode component $\alpha$ is raised to 0.46, while the coherent 
multimode component $\beta$ drops to 0.34. Thus the single mode 
contribution is increased because 
the value of $r>0$ observed in the experiment is now partially 
attributed to the effects of decoherence, reducing the 
multi--mode contribution necessary to explain the
experimentally observed correlations.

A more precise investigation of decoherence effects will be 
presented elsewhere. Here, it should only be noted that 
decoherence effects generally increase the ratio $r$, since 
decoherence reduces the polarization correlations that are 
responsible for the differences between $C_\parallel$ and 
$C_\perp$. The minimal fraction of $ETP$ emission required to 
explain the observed value of $r<1/2$ is therefore obtained
by assuming the emission of pure state $ETP$ and $EOP$, as 
given in Eqs.~(8)--(11). 
Our method has thus successfully verified the presence 
of a significant single mode component in the emission
\cite{footnote2}.

In conclusion, a scheme to distinguish $ETP$ from two independent 
$EOP$s based on four--fold coincidence has been 
proposed and demonstrated experimentally. The experimental results 
indicate the 
generation of two photons in the same spatiotemporal mode in 
each path, strongly suggesting the formation of $ETP$. 
Specifically, we have been able to confirm a minimum of 37$\%$ single 
mode emission among the entangled
four photon states generated in our experimental setup.

We would like to thank J. Hotta, H. Fujiwara, H. Oka, A. G. White and 
F. Morikoshi for their useful discussions. 
We also thank D. Kawase, K. Ushizaka and T. Tsujioka for technical support. 
This work was supported 
by Core Research for Evolutional Science and Technology, 
Japan Science and Technology Agency. Part of this work was 
also supported by the International Communication Foundation.
%
%

%

\begin{thebibliography}{1}
\bibitem{footnote}
More precisely, it can be shown that this assumption is generally
valid if there is no local polarization, a situation that 
applies to the present experiment.
\bibitem{footnote2}
It should be noted that the fraction of single mode emission 
is not related to the amount of entanglement
generated, since the $ETP$ fraction is generally mixed state 
$ETP$ with non--maximal entanglement.
\end{thebibliography}
\end{document}